\documentclass[preprints,article,accept,moreauthors,pdftex,10pt,a4paper]{Definitions/mdpi}

\pdfoutput=1

\firstpage{1}
\pubvolume{xx}
\issuenum{1}
\articlenumber{1}
\pubyear{2018}
\copyrightyear{2018}
\externaleditor{}
\history{}

\usepackage{mathptmx}
\usepackage{mathrsfs}
\usepackage{bbold}
\usepackage{amssymb}
\usepackage{bm}
\usepackage[LY1]{fontenc}
\newcommand\ket[1]{\,|#1\rangle}

\newcommand\braket[2]{\langle #1 | #2\rangle}
\newcommand\Span[0]{\operatorname{Span}}

\newcommand\bket[1]{|\mathfrak{#1}\rangle}

\newcommand\bitz[0]{\mathfrak{0}}
\newcommand\bito[0]{\mathfrak{1}}
\newcommand\sA[0]{\mathrm{A}}
\newcommand\sB[0]{\mathrm{B}}
\newcommand\sE[0]{\mathrm{E}}
\newcommand\sx[0]{\mathrm{x}}
\newcommand\sxB[0]{\mathrm{x,B}}
\newcommand\sCTRL[0]{\mathrm{CTRL}}
\newcommand\sSWAPx[0]{\mathrm{SWAP\text{-}x}}

\Title{Attacks against a Simplified Experimentally Feasible
Semiquantum Key Distribution Protocol}

\Author{Michel Boyer $^{1}$, Rotem Liss $^{2,}$*\orcidA{}
and Tal Mor $^{2}$}

\AuthorNames{Michel Boyer, Rotem Liss and Tal Mor}

\address{$^{1}$ \quad D\'epartement d'informatique et de recherche
op\'erationnelle (DIRO), Universit\'e de Montr\'eal,
Montr\'eal, QC H3C 3J7, Canada; boyer@iro.umontreal.ca \\
$^{2}$ \quad Computer Science Department, Technion, Haifa 3200003,
Israel; rotemliss@cs.technion.ac.il; talmo@cs.technion.ac.il}

\corres{Correspondence: rotemliss@cs.technion.ac.il}

\abstract{A semiquantum key distribution (SQKD) protocol makes it
possible for a quantum party and a classical party
to generate a secret shared key.
However, many existing SQKD protocols are not experimentally feasible
in a secure way using current technology.
An experimentally feasible SQKD protocol,
``classical Alice with a controllable mirror''
(the ``Mirror protocol''), has recently been presented
and proved completely robust,
but it is more complicated than other SQKD protocols.
Here we prove a simpler variant of the Mirror protocol
(the ``simplified Mirror protocol'') to be completely non-robust
by presenting two possible attacks against it.
Our results show that the complexity of the Mirror protocol
is at least partly necessary for achieving robustness.}


\begin{document}

\section{Introduction}
Quantum key distribution (QKD) protocols allow
two parties, Alice and Bob,
to share a secret random key that is
secure even against the most powerful adversaries.
Semiquantum key distribution (SQKD) protocols achieve the same goal
even if one of the two parties (Alice or Bob)
is limited to use only classical operations:
the classical party can use only the computational basis
$\{\bket{0}, \bket{1}\}$, while the quantum party can use any basis --
for example, both the computational basis and the Hadamard basis
$\{\bket{\text{+}} \triangleq \frac{\bket{0} + \bket{1}}{\sqrt{2}},
\bket{-} \triangleq \frac{\bket{0} - \bket{1}}{\sqrt{2}}\}$.
As explained in~\cite{cbob07,sqkd09},
the importance of SQKD protocols is both conceptual and practical:
they make it possible to investigate the amount of ``quantumness''
needed for QKD, and they may, in some cases, be easier to implement
than standard QKD protocols.

The first SQKD protocol was ``QKD with classical Bob''~\cite{cbob07}.
Later, other SQKD protocols have been suggested, including
``QKD with classical Alice''~\cite{calice09,calice09comment}
and many others
(e.g.,~\cite{sqkd09,LC2008,SDL2013,YYLH2014,mediated15,ZQZM15}).
Most SQKD protocols have been proven ``robust'':
namely~\cite{cbob07}, if the adversary Eve succeeds
in getting some secret information,
she must cause some errors that may be noticed by Alice and Bob.
A few SQKD protocols also have a security
analysis~\cite{cbob_secur15,sqkd_secur16,calice_secur18,sqkd_secur18}.
Proving robustness is a step towards proving security;
proving the security of SQKD protocols is difficult
because those protocols are usually two-way: for example, Alice sends
a quantum state to Bob, and Bob performs a specific classical operation
and sends the resulting quantum state back to Alice.

However, many SQKD protocols, including~\cite{cbob07,calice09},
are vulnerable to practical attacks and cannot be experimentally
constructed in a secure way using current technology.
An important classical operation of those protocols is named SIFT.
The definition of a SIFT operation performed by Alice
(assuming that Alice is the classical party) is as follows:
Alice measures the incoming quantum state
in the computational basis $\{\bket{0}, \bket{1}\}$
and then generates the state she measured and resends it towards Bob.
Security of those SQKD protocols relies on the assumption
that during the SIFT operation, Alice's measurement devices
can measure the \emph{precise} states $\{\bket{0}, \bket{1}\}$
and distinguish those precise states from any imperfect similar state,
and Alice's photon generation devices can generate
the \emph{precise} states $\{\bket{0}, \bket{1}\}$
and not any other (imperfect) state.
In particular, the generated states $\{\bket{0}, \bket{1}\}$
must be indistinguishable from states that
Alice \emph{reflects} towards Bob.
Using current photonic technology, Alice's devices are imperfect,
which makes this assumption incorrect and
makes possible attacks by the eavesdropper Eve:
for example, Eve may send a slightly modified state towards Alice
(a ``tagging attack'')
or may distinguish between the states sent by Alice.
Full details about those practical attacks are available
in~\cite{cbob07comment,cbob07comment_reply,mirror17}.

An experimentally feasible SQKD protocol named
``classical Alice with a controllable mirror''
(the ``Mirror protocol'') has recently been presented~\cite{mirror17}.
This protocol is safe against the
``tagging'' attack presented by~\cite{cbob07comment}.
Moreover, the protocol was proved by~\cite{mirror17}
to be completely robust against any attacker Eve,
even if Eve is all-powerful and limited only by the laws of physics,
and even if Eve can send multi-photon pulses.
The robustness proof is still correct even if the detectors of
Alice and Bob cannot \emph{count} how many photons arrive in each mode:
namely, when either Alice or Bob looks at a detector which detects
a specific mode, they can only notice whether it ``clicks''
(detects one photon or more in that mode)
or not (finds the mode to be empty).
This is the standard situation when using current technology.

In this paper, we present a simpler variant of the Mirror protocol
(the ``simplified Mirror protocol'') which is easier to implement.
Our variant allows the classical party, Alice, to choose one of
three operations, while the Mirror protocol allows Alice
to choose one of four operations.
We present two attacks against this variant,
proving it to be non-robust.
Our results show that the four classical operations allowed by
the Mirror protocol are probably necessary for robustness.

In Section~\ref{sec_mirror} we present the Mirror protocol
described by~\cite{mirror17}. In Section~\ref{sec_simplified} we present
the simplified Mirror protocol and its motivation.
In Section~\ref{sec_attacks} we prove the simplified Mirror protocol
to be non-robust by presenting two attacks against it:
a full attack and a weaker attack. In Section~\ref{sec_discuss}
we discuss potential implications of our results.

\section{\label{sec_mirror}The Mirror Protocol}
For describing the Mirror protocol (presented by~\cite{mirror17}),
we assume a photonic implementation consisting of two modes:
the mode of the qubit state $\bket{0}$
and the mode of the qubit state $\bket{1}$ (below we call them
``the $\bket{0}$ mode'' and ``the $\bket{1}$ mode'', respectively).
For example, the $\bket{0}$ mode and the $\bket{1}$ mode
can represent two different polarizations or two different time bins.
We use the Fock space notations: if there is exactly one photon
(and, thus, our Hilbert space is the qubit space),
the Fock state $\ket{0,1}$ (equivalent to $\ket{0}$)
represents one photon in the $\bket{0}$ mode,
and the Fock state $\ket{1,0}$ (equivalent to $\ket{1}$)
represents one photon in the $\bket{1}$ mode.
We can extend the qubit space to a $3$-dimensional Hilbert space
by adding the Fock ``vacuum state'' $\ket{0,0}$,
which represents an absence of photons.
Most generally, the Fock state $\ket{m_\bito, m_\bitz}$ represents
$m_\bito$ indistinguishable photons in the $\bket{1}$ mode
and $m_\bitz$ indistinguishable photons in the $\bket{0}$ mode.
Similarly (in the Hadamard basis), the Fock state $\ket{m_-, m_+}_\sx$
represents $m_-$ indistinguishable photons in the $\bket{-}$ mode
and $m_+$ indistinguishable photons in the $\bket{\text{+}}$ mode.
More details about the Fock space notations are given
in~\cite{mirror17}; it is vital to use those mathematical notations
for describing and analyzing all practical attacks
on a QKD protocol (see~\cite{BLMS00} for details).

In the Mirror protocol, in each round,
Bob sends to Alice the $\ket{\text{+}}_\sB$ state --
namely, the $\ket{\text{0,1}}_\sxB \triangleq
\frac{\ket{0,1}_\sB + \ket{1,0}_\sB}{\sqrt{2}}$ state.
Then, Alice prepares an ancillary state in the initial vacuum state
$\ket{0,0}_\sA$ and chooses at random one of the following
four classical operations:
\begin{description}
\item[$\mathbf{I}$ (CTRL)] Reflect all the photons towards Bob,
without measuring any photon. The mathematical description is:
\begin{equation}
I \ket{0,0}_\sA \ket{m_\bito,m_\bitz}_\sB
= \ket{0,0}_\sA \ket{m_\bito,m_\bitz}_\sB.
\end{equation}
\item[$\mathbf{S_1}$ (SWAP-10)] Reflect all photons in the
$\bket{0}$ mode towards Bob, and measure all photons in the
$\bket{1}$ mode. The mathematical description is:
\begin{equation}
S_1 \ket{0,0}_\sA \ket{m_\bito,m_\bitz}_\sB
= \ket{{m_\bito,0}}_\sA \ket{0,m_\bitz}_\sB.
\end{equation}
\item[$\mathbf{S_0}$ (SWAP-01)] Reflect all photons in the
$\bket{1}$ mode towards Bob, and measure all photons in the
$\bket{0}$ mode. The mathematical description is:
\begin{equation}
S_0 \ket{0,0}_\sA \ket{m_\bito,m_\bitz}_\sB
= \ket{0,m_\bitz}_\sA \ket{m_\bito,0}_\sB.
\end{equation}
\item[$\mathbf{S}$ (SWAP-ALL)] Measure all the photons, without
reflecting any photon towards Bob. The mathematical description is:
\begin{equation}
S \ket{0,0}_\sA \ket{m_\bito,m_\bitz}_\sB
= \ket{m_\bito,m_\bitz}_\sA \ket{0,0}_\sB.
\end{equation}
\end{description}
(We note that in the above mathematical description, Alice measures
her ancillary state $\ket{\cdot}_\sA$ in the computational basis
and sends back to Bob the $\ket{\cdot}_\sB$ state.)

The states sent from Alice to Bob (without any error, loss,
or eavesdropping) are detailed in Table~\ref{table_expected_results}.
\begin{table}[H]
\caption{The state sent from Alice to Bob in the Mirror protocol
without errors or losses, depending on Alice's classical operation and
on whether Alice detected a photon or not.}
\label{table_expected_results}
\centering
\begin{tabular}{ccc}
\toprule
\textbf{Alice's Classical Operation} &
\textbf{Did Alice Detect a Photon?} &
\textbf{State Sent from Alice to Bob} \\
\midrule
CTRL & no (happens with certainty) & $\ket{0,1}_\sxB = \frac{1}{\sqrt{2}} \left[ \ket{0,1}_\sB + \ket{1,0}_\sB \right]$ \\
\midrule
SWAP-10 & no (happens with probability $\frac{1}{2}$) &
$\ket{0,1}_\sB$ \\
SWAP-10 & yes (happens with probability $\frac{1}{2}$) &
$\ket{0,0}_\sB$ \\
\midrule
SWAP-01 & no (happens with probability $\frac{1}{2}$) &
$\ket{1,0}_\sB$ \\
SWAP-01 & yes (happens with probability $\frac{1}{2}$) &
$\ket{0,0}_\sB$ \\
\midrule
SWAP-ALL & yes (happens with certainty) & $\ket{0,0}_\sB$ \\
\bottomrule
\end{tabular}
\end{table}

Then, Bob measures the incoming state in a random basis
(either the computational basis $\{\ket{0}, \ket{1}\}$ or
the Hadamard basis $\{\ket{\text{+}}, \ket{-}\}$).
After completing all rounds, Alice sends over the classical
channel her operation choices (CTRL, SWAP-$x$, or SWAP-ALL;
she keeps $x \in \{01,10\}$ in secret),
Bob sends over the classical channel his basis choices,
and both of them reveal some non-secret information
on their measurement results (as elaborated in~\cite{mirror17}).
Then, Alice and Bob reveal and compute the error rate
on test bits for which Alice used SWAP-10 or SWAP-01 and Bob measured
in the computational basis, and the error rate on test bits for which
Alice used CTRL and Bob measured in the Hadamard basis.
They also check whether other errors exist (for example,
they verify Bob detects no photons in case Alice uses SWAP-ALL).
Alice and Bob also discard mismatched rounds, such as rounds in which
Alice used SWAP-10 and Bob used the Hadamard basis.
Alice and Bob share the secret bit 0 if Alice uses
SWAP-10 and detects no photon while Bob measures in the
computational basis and detects a photon in the $\bket{0}$ mode;
similarly, they share the secret bit 1 if Alice uses
SWAP-01 and detects no photon while Bob measures in the
computational basis and detects a photon in the $\bket{1}$ mode.

Finally, Alice and Bob verify that the error rates are below some
thresholds, and they perform error correction and privacy amplification
in the standard way for QKD protocols.
At the end of the protocol, Alice and Bob hold an identical final key
that is completely secure against any eavesdropper.

A full description of the protocol and a proof of its complete
robustness are both available in~\cite{mirror17}.

The experimental implementation of the protocol
can use two time bins (namely, two pulses),
one for the $\bket{0}$ mode and one for the $\bket{1}$ mode.
In this case, Alice's possible operations can be described
as possible ways for operating a controllable mirror, so that Alice can
choose whether to reflect or measure the photon(s) in each time bin.
The mirror can be experimentally implemented in various ways;
for example:
\begin{itemize}
\item It can be implemented as a mechanically moved mirror.
Such mirror is trivial to implement, but it is very slow.
\item It can be implemented by using optical elements: an
electronically-triggered Pockels cell, which changes the polarization
of the photon(s) in one of the pulses, and a polarizing beam splitter,
which can split the two different pulses (that now have different
polarizations) into two paths. This implementation is feasible and
gives much higher bit rates than the mechanical implementation.
\end{itemize}
More details about the experimental implementations
are available in~\cite{mirror17}.

\section{\label{sec_simplified}The ``Simplified Mirror Protocol'':
a Simpler and Non-Robust Variant of the Mirror Protocol}
In this paper, we discuss a simpler variant of the Mirror protocol,
which we name the ``simplified Mirror protocol''.
The simplified Mirror protocol is identical to
the Mirror protocol described in Section~\ref{sec_mirror},
except that it does not include the SWAP-ALL operation.
In other words, in the simplified protocol, Alice chooses at random
one of the three classical operations CTRL, SWAP-10, and SWAP-01.

The simplified protocol is easier to implement, because the SWAP-ALL
operation poses some experimental challenges to the electronic
implementation discussed in Section~\ref{sec_mirror}:
for implementing SWAP-ALL, the Pockels cell should either remain working
for a long time (changing polarization for both time bins) or
be operated twice (changing polarization for each time bin separately).
In more details, for the two pulses
representing the $\bket{0}$ mode and the $\bket{1}$ mode:
if we assume the duration of each pulse is $t$ and
the time difference between the two pulses is $T$ (where $t \ll T$),
the first solution means keeping the Pockels cell
operating during the time period $[0,T+2t]$,
and the second solution means operating the Pockels cell
during the two time periods $[0,t]$ and $[T+t,T+2t]$.
The first solution may be problematic for some models of the Pockels
cell, and the second solution may be problematic
because of the recovery time needed for the Pockels cell.
Therefore, at least in some implementations,
the simplified Mirror protocol is much easier to implement
than the standard Mirror protocol.

Moreover, analyzing the simplified protocol gives a better
understanding of the properties required for an SQKD protocol to be
robust. In particular, this analysis explains why the structure and
complexity of the Mirror protocol are necessary for robustness.

For completeness, we provide below the full description of
the simplified Mirror protocol. We note that this description is almost
the same as the description of the Mirror protocol
in Section~\ref{sec_mirror}.

In the simplified Mirror protocol, in each round,
Bob sends to Alice the $\ket{\text{+}}_\sB$ state --
namely, the $\ket{\text{0,1}}_\sxB \triangleq
\frac{\ket{0,1}_\sB + \ket{1,0}_\sB}{\sqrt{2}}$ state.
Then, Alice prepares an ancillary state in the initial vacuum state
$\ket{0,0}_\sA$ and chooses at random one of the following
three classical operations:
\begin{description}
\item[$\mathbf{I}$ (CTRL)] Reflect all the photons towards Bob,
without measuring any photon. The mathematical description is:
\begin{equation}
I \ket{0,0}_\sA \ket{m_\bito,m_\bitz}_\sB
= \ket{0,0}_\sA \ket{m_\bito,m_\bitz}_\sB.
\end{equation}
\item[$\mathbf{S_1}$ (SWAP-10)] Reflect all photons in the
$\bket{0}$ mode towards Bob, and measure all photons in the
$\bket{1}$ mode. The mathematical description is:
\begin{equation}
S_1 \ket{0,0}_\sA \ket{m_\bito,m_\bitz}_\sB
= \ket{{m_\bito,0}}_\sA \ket{0,m_\bitz}_\sB.
\end{equation}
\item[$\mathbf{S_0}$ (SWAP-01)] Reflect all photons in the
$\bket{1}$ mode towards Bob, and measure all photons in the
$\bket{0}$ mode. The mathematical description is:
\begin{equation}
S_0 \ket{0,0}_\sA \ket{m_\bito,m_\bitz}_\sB
= \ket{0,m_\bitz}_\sA \ket{m_\bito,0}_\sB.
\end{equation}
\end{description}
(We note that in the above mathematical description, Alice measures
her ancillary state $\ket{\cdot}_\sA$ in the computational basis
and sends back to Bob the $\ket{\cdot}_\sB$ state.)

The states sent from Alice to Bob
(without any error, loss, or eavesdropping)
are detailed in Table~\ref{table_simp_expected_results}.
\begin{table}[H]
\caption{The state sent from Alice to Bob
in the simplified Mirror protocol without errors or losses,
depending on Alice's classical operation and
on whether Alice detected a photon or not.}
\label{table_simp_expected_results}
\centering
\begin{tabular}{ccc}
\toprule
\textbf{Alice's Classical Operation} &
\textbf{Did Alice Detect a Photon?} &
\textbf{State Sent from Alice to Bob} \\
\midrule
CTRL & no (happens with certainty) & $\ket{0,1}_\sxB = \frac{1}{\sqrt{2}} \left[ \ket{0,1}_\sB + \ket{1,0}_\sB \right]$ \\
\midrule
SWAP-10 & no (happens with probability $\frac{1}{2}$) &
$\ket{0,1}_\sB$ \\
SWAP-10 & yes (happens with probability $\frac{1}{2}$) &
$\ket{0,0}_\sB$ \\
\midrule
SWAP-01 & no (happens with probability $\frac{1}{2}$) &
$\ket{1,0}_\sB$ \\
SWAP-01 & yes (happens with probability $\frac{1}{2}$) &
$\ket{0,0}_\sB$ \\
\bottomrule
\end{tabular}
\end{table}

Then, Bob measures the incoming state in a random basis
(either the computational basis $\{\ket{0}, \ket{1}\}$ or
the Hadamard basis $\{\ket{\text{+}}, \ket{-}\}$).
After completing all rounds, Alice sends over the classical
channel her operation choices (CTRL or SWAP-$x$;
she keeps $x \in \{01,10\}$ in secret),
Bob sends over the classical channel his basis choices,
and both of them reveal some non-secret information
on their measurement results (as elaborated in~\cite{mirror17}).
Then, Alice and Bob reveal and compute the error rate
on test bits for which Alice used SWAP-10 or SWAP-01 and Bob measured
in the computational basis, and the error rate on test bits for which
Alice used CTRL and Bob measured in the Hadamard basis.
They also check whether other errors exist (for example,
it must never happen that \emph{both} Alice and Bob detect a photon).
Alice and Bob also discard mismatched rounds, such as rounds in which
Alice used SWAP-10 and Bob used the Hadamard basis.
Alice and Bob share the secret bit 0 if Alice uses
SWAP-10 and detects no photon while Bob measures in the
computational basis and detects a photon in the $\bket{0}$ mode;
similarly, they share the secret bit 1 if Alice uses
SWAP-01 and detects no photon while Bob measures in the
computational basis and detects a photon in the $\bket{1}$ mode.

Finally, Alice and Bob verify that the error rates are below some
thresholds, and they perform error correction and privacy amplification
in the standard way for QKD protocols.
At the end of the protocol, Alice and Bob hold an identical final key
that is completely secure against any eavesdropper.

\section{\label{sec_attacks}Attacks against
the Simplified Mirror Protocol}
We prove the simplified protocol to be non-robust
by presenting two attacks:
a ``full attack'' described in Subsection~\ref{subsec_attack_full},
which gives Eve full information
but causes full loss of the CTRL bits,
and a ``weaker attack'' described
in Subsection~\ref{subsec_attack_weaker},
which gives Eve less information
but causes fewer losses of CTRL bits.

\subsection{\label{subsec_attack_full}A Full Attack on the Simplified
Protocol that Gives Eve Full Information}
In this attack, Eve gets full information of all the information bits.
Namely, she gets full information on the SWAP-10 and SWAP-01 bits that
were measured by Bob in the computational basis.

Eve applies her attack in two stages:
the first stage is on the way from Bob to Alice,
and the second stage is on the way from Alice to Bob.
In both stages she uses her own probe space (namely, ancillary space)
$\mathscr{H}_\sE = \mathscr{H}_3$ spanned by the orthonormal basis
$\{\ket{0}_\sE, \ket{1}_\sE, \ket{2}_\sE\}$.
We assume that Eve fully controls the environment,
the errors, and the losses (this is a standard assumption
when analyzing the security of QKD):
namely, no losses and no errors exist between Bob and Eve
or between Alice and Eve.

In the first stage of the attack (on the way from Bob to Alice),
Eve intercepts the state $\ket{\text{+}}_\sB$
(namely, $\ket{0,1}_\sxB$) sent by Bob, generates instead the state
\begin{equation}
\frac{1}{\sqrt{3}} \left[ \ket{0,1}_\sB \ket{1}_\sE
+ \ket{1,0}_\sB \ket{1}_\sE + \ket{0,0}_\sB \ket{0}_\sE \right]
= \sqrt{\frac{2}{3}} \ket{0,1}_\sxB \ket{1}_\sE
+ \sqrt{\frac{1}{3}} \ket{0,0}_\sB \ket{0}_\sE,\label{simplified_attack}
\end{equation}
and sends to Alice the $\sB$ part of the state.
This state causes Alice to get no photons with probability
$\frac{1}{3}$ and get the expected $\ket{\text{+}}_\sB$ state
with probability $\frac{2}{3}$.
Alice then performs at random one of the three classical operations
CTRL, SWAP-10, or SWAP-01. The resulting possible states of Bob+Eve
are described in Table~\ref{table_AE_results}.

\begin{table}[H]
\caption{The state of Bob+Eve after Alice's classical operation
for the attacks described in Subsections~\ref{subsec_attack_full}
and~\ref{subsec_attack_weaker}, depending on Alice's classical
operation and on whether Alice detected a photon or not.}
\label{table_AE_results}
\centering
\begin{tabular}{ccc}
\toprule
\textbf{Alice's Classical Operation} &
\textbf{Did Alice Detect a Photon?} & \textbf{Bob+Eve State} \\
\midrule
CTRL & no (happens with certainty) & $\frac{1}{\sqrt{3}}
\left[ \ket{0,1}_\sB \ket{1}_\sE + \ket{1,0}_\sB \ket{1}_\sE
+ \ket{0,0}_\sB \ket{0}_\sE \right]$ \\
\midrule
SWAP-10 & no (happens with probability $\frac{2}{3}$) &
$\frac{1}{\sqrt{2}} \left[ \ket{0,1}_\sB \ket{1}_\sE
+ \ket{0,0}_\sB \ket{0}_\sE \right]$ \\
SWAP-10 & yes (happens with probability $\frac{1}{3}$) &
$\ket{0,0}_\sB \ket{1}_\sE$ \\
\midrule
SWAP-01 & no (happens with probability $\frac{2}{3}$) &
$\frac{1}{\sqrt{2}} \left[ \ket{1,0}_\sB \ket{1}_\sE
+ \ket{0,0}_\sB \ket{0}_\sE \right]$ \\
SWAP-01 & yes (happens with probability $\frac{1}{3}$) &
$\ket{0,0}_\sB \ket{1}_\sE$ \\
\bottomrule
\end{tabular}
\end{table}

In the second stage of the attack (on the way from Alice to Bob),
Eve applies the unitary operator $V$ on the joint Bob+Eve state,
where $V$ is defined as follows:
\begin{eqnarray}
V \ket{0,1}_\sB \ket{1}_\sE
&=& -\sqrt{\frac{1}{3}} \ket{1,0}_\sB \ket{1}_\sE
+ \sqrt{\frac{2}{3}} \ket{0,0}_\sB \ket{0}_\sE, \\
V \ket{1,0}_\sB \ket{1}_\sE
&=& -\sqrt{\frac{1}{3}} \ket{0,1}_\sB \ket{0}_\sE
+ \sqrt{\frac{2}{3}} \ket{0,0}_\sB \ket{1}_\sE, \\
V \ket{0,0}_\sB \ket{0}_\sE
&=& \sqrt{\frac{1}{3}} \ket{0,1}_\sB \ket{0}_\sE
+ \sqrt{\frac{1}{3}} \ket{1,0}_\sB \ket{1}_\sE
+ \sqrt{\frac{1}{3}} \ket{0,0}_\sB \ket{\text{+}}_\sE, \\
V \ket{0,0}_\sB \ket{1}_\sE &=& \ket{0,0}_\sB \ket{2}_\sE.
\end{eqnarray}

$V$ is indeed a unitary operator, because we can prove
the right-hand sides to be orthonormal: all the right-hand sides
are normalized vectors; the first two vectors are clearly orthogonal;
the third vector is orthogonal to the first two, because
$\braket{0}{\text{+}}_\sE = \braket{1}{\text{+}}_\sE
= \frac{1}{\sqrt{2}}$;
and the fourth vector is orthogonal to the three others.
Thus, $V$ defines (or, more precisely, can be extended to) a unitary
operator on $\mathscr{H}_\sB \otimes \mathscr{H}_\sE$.

Applying the unitary operator $V$ on Table~\ref{table_AE_results}
gives the states listed in Table~\ref{table_BE_results_full}.
Comparing it with Table~\ref{table_simp_expected_results},
we conclude that this attack never causes Alice and Bob
to detect an error. Moreover, Eve detects the whole secret key:
Eve measures ``$0$'' in her probe if Alice and Bob agree
on the bit $0$, and she measures ``$1$'' in her probe if Alice and Bob
agree on the bit $1$. However, Eve causes several kinds of losses;
in particular, all the CTRL bits are lost.

\begin{table}[H]
\caption{The state of Bob+Eve after completing Eve's attack
described in Subsection~\ref{subsec_attack_full}, depending on Alice's
classical operation and on whether Alice detected a photon or not.}
\label{table_BE_results_full}
\centering
\begin{tabular}{ccc}
\toprule
\textbf{Alice's Classical Operation} &
\textbf{Did Alice Detect a Photon?} & \textbf{Bob+Eve State} \\
\midrule
CTRL & no (happens with certainty) &
$\ket{0,0}_\sB \ket{\text{+}}_\sE$ \\
\midrule
SWAP-10 & no (happens with probability $\frac{2}{3}$) &
$\displaystyle\frac{1}{\sqrt{6}} \ket{0,1}_\sB \ket{0}_\sE
+ \ket{0,0}_\sB \frac{3 \ket{0}_\sE + \ket{1}_\sE}{\sqrt{12}}$ \\
SWAP-10 & yes (happens with probability $\frac{1}{3}$) &
$\ket{0,0}_\sB \ket{2}_\sE$ \\
\midrule
SWAP-01 & no (happens with probability $\frac{2}{3}$) &
$\displaystyle\frac{1}{\sqrt{6}} \ket{1,0}_\sB \ket{1}_\sE
+ \ket{0,0}_\sB \frac{\ket{0}_\sE + 3\ket{1}_\sE}{\sqrt{12}}$ \\
SWAP-01 & yes (happens with probability $\frac{1}{3}$) &
$\ket{0,0}_\sB \ket{2}_\sE$ \\
\bottomrule
\end{tabular}
\end{table}

Therefore, this attack makes it possible for Eve to get full information
without inducing any error. However, Eve causes many losses, including
full loss of the CTRL bits.

\subsection{\label{subsec_attack_weaker}A Weaker Attack
on the Simplified Protocol Causing Fewer Losses of the CTRL Bits}
The full attack described in Subsection~\ref{subsec_attack_full}
makes it impossible for Bob to ever detect a CTRL bit,
which may look suspicious.
We now present a weaker attack that lets Bob detect
some CTRL bits but gives Eve less information.

The first stage of the attack (on the way from Bob to Alice)
remains the same: that is, the state Eve sends to Alice
is still given by Equation~\eqref{simplified_attack},
and the resulting Bob+Eve state after Alice's classical operation
is still shown in Table~\ref{table_AE_results}.
Eve's probe space is, too, the same as before:
$\mathscr{H}_\sE = \mathscr{H}_3 \triangleq
\Span \{\ket{0}_\sE, \ket{1}_\sE, \ket{2}_\sE\}$.

This attack is characterized by the parameter $0 \le \epsilon \le 1$.
We will see that $\epsilon = 0$ gives the full attack
described in Subsection~\ref{subsec_attack_full},
while $\epsilon = 1$ gives Eve no information at all.

Another important parameter used by the attack is
\begin{equation}
\kappa \triangleq \sqrt{\frac{1 - \epsilon^2}{3 - 2\epsilon^2}}.
\end{equation}

We notice that for small values of $\epsilon$,
the value of $\kappa$ is close to $\sqrt{\frac{1}{3}}$.
Moreover, for all $0 \le \epsilon \le 1$, it holds that
$0 < \epsilon^2 + \kappa^2 \le 1$ and $2\kappa^2 < 1$.

In the second stage of the attack (on the way from Alice to Bob),
Eve applies the unitary operator $V$ on the joint Bob+Eve state,
where $V$ is defined as follows:
\begin{eqnarray}
V \ket{0,1}_\sB \ket{1}_\sE
&=& \phantom{+} \epsilon \ket{0,1}_\sB \ket{2}_\sE
- \kappa \ket{1,0}_\sB \ket{1}_\sE
+ \sqrt{1 - \kappa^2 - \epsilon^2} \ket{0,0}_\sB \ket{0}_\sE, \\
V \ket{1,0}_\sB \ket{1}_\sE
&=& -\kappa \ket{0,1}_\sB \ket{0}_\sE
+ \epsilon \ket{1,0}_\sB \ket{2}_\sE
+ \sqrt{1 - \kappa^2 - \epsilon^2} \ket{0,0}_\sB \ket{1}_\sE, \\
V \ket{0,0}_\sB \ket{0}_\sE
&=& \phantom{+} \kappa \ket{0,1}_\sB \ket{0}_\sE
+ \kappa \ket{1,0}_\sB \ket{1}_\sE
+ \sqrt{1 - 2\kappa^2} \ket{0,0}_\sB \ket{\text{+}}_\sE, \\
V \ket{0,0}_\sB \ket{1}_\sE &=& \phantom{+} \ket{0,0}_\sB \ket{2}_\sE.
\end{eqnarray}

$V$ is indeed a unitary operator, because we can prove
the right-hand sides to be orthonormal: all the right-hand
sides are clearly normalized; the first two vectors are orthogonal;
the fourth vector is orthogonal to the three others;
and the third vector is orthogonal to the first and to the second,
because
\begin{eqnarray}
1 - 2\kappa^2 &=& \frac{3 - 2\epsilon^2 - 2 (1 - \epsilon^2)}
{3 - 2\epsilon^2} = \frac{1}{3 - 2\epsilon^2}, \\
1 - \kappa^2 - \epsilon^2 &=& \frac{(3 - 2\epsilon^2)
- (1 - \epsilon^2) - (3\epsilon^2 - 2\epsilon^4)}{3 - 2\epsilon^2}
= \frac{2(1 - \epsilon^2)^2}{3 - 2\epsilon^2},
\end{eqnarray}
and thus $\frac{\sqrt{1 - \kappa^2 - \epsilon^2} \sqrt{1 - 2\kappa^2}}
{\sqrt{2}} = \kappa^2$.
Therefore, $V$ extends to a unitary operator on
$\mathscr{H}_\sB \otimes \mathscr{H}_\sE$.

The final global state after Eve's attack is described in
Table~\ref{table_BE_results_weaker} (calculated by applying
the operator $V$ on Table~\ref{table_AE_results}),
given the following definitions:
\begin{eqnarray}
a &\triangleq& \sqrt{1 - \kappa^2 - \epsilon^2}
+ \frac{\sqrt{1 - 2\kappa^2}} {\sqrt{2}},\label{weaker_a_def} \\
b &\triangleq& \frac{\sqrt{1 - 2\kappa^2}}{\sqrt{2}}.
\label{weaker_b_def}
\end{eqnarray}

\begin{table}[H]
\caption{The state of Bob+Eve after completing Eve's attack
described in Subsection~\ref{subsec_attack_weaker},
depending on Alice's classical operation
and on whether Alice detected a photon or not.
The parameters $a$ and $b$ are defined
in Equations~\eqref{weaker_a_def} and~\eqref{weaker_b_def}.}
\label{table_BE_results_weaker}
\centering
\begin{tabular}{ccc}
\toprule
\textbf{Alice's Classical Operation} &
\textbf{Did Alice Detect a Photon?} & \textbf{Bob+Eve State} \\
\midrule
CTRL & no (happens with certainty) &
$\displaystyle\sqrt{\frac{2 \epsilon^2}{3}} \ket{0,1}_\sxB \ket{2}_\sE
+ \sqrt{1 - \frac{2 \epsilon^2}{3}} \ket{0,0}_\sB \ket{\text{+}}_\sE$ \\
\midrule
SWAP-10 & no (happens with probability $\frac{2}{3}$) &
$\displaystyle\frac{1}{\sqrt{2}} \left[ \ket{0,1}_\sB
\left( \epsilon \ket{2}_\sE + \kappa \ket{0}_\sE \right)
+ \ket{0,0}_\sB \left( a \ket{0}_\sE + b \ket{1}_\sE \right) \right]$
\\
SWAP-10 & yes (happens with probability $\frac{1}{3}$) &
$\displaystyle\ket{0,0}_\sB \ket{2}_\sE$ \\
\midrule
SWAP-01 & no (happens with probability $\frac{2}{3}$) &
$\displaystyle\frac{1}{\sqrt{2}} \left[ \ket{1,0}_\sB
\left( \epsilon \ket{2}_\sE + \kappa \ket{1}_\sE \right)
+ \ket{0,0}_\sB \left( b \ket{0}_\sE + a\ket{1}_\sE \right) \right]$
\\
SWAP-01 & yes (happens with probability $\frac{1}{3}$) &
$\displaystyle\ket{0,0}_\sB \ket{2}_\sE$ \\
\bottomrule
\end{tabular}
\end{table}

We notice that for $\epsilon = 0$, the attack is the same as in
Subsection~\ref{subsec_attack_full}.
If $\epsilon = 1$, the loss rate of CTRL bits is
$\frac{1}{3}$, and Eve gets no information at all
on the information bits (because $\kappa = 0$).

In general, if Alice and Bob share a ``secret'' bit $b \in \{0,1\}$,
Eve's probe state is in the (normalized) state
\begin{equation}
\frac{\epsilon \ket{2}_\sE + \kappa \ket{b}_\sE}
{\sqrt{\epsilon^2 + \kappa^2}}.
\end{equation}

When Eve measures her probe state in the computational basis
$\{\ket{0}_\sE, \ket{1}_\sE, \ket{2}_\sE\}$,
she gets the information bit $b$ with probability
\begin{equation}
p = \frac{\kappa^2}{\epsilon^2 + \kappa^2}
= \frac{1 - \epsilon^2}{1 + 2\epsilon^2 - 2\epsilon^4},
\end{equation}
and the loss rates of CTRL and SWAP-$x$ bits
(where $x \in \{01,10\}$) are
\begin{eqnarray}
R_\sCTRL &=& 1 - \frac{2 \epsilon^2}{3}, \\
R_\sSWAPx &=& 1 - \frac{\epsilon^2 + \kappa^2}{2},
\end{eqnarray}
respectively.

Table~\ref{table_epsilon_p_weaker} shows the probabilities $p$ and
the loss rates $R_\sCTRL,R_\sSWAPx$ for various values of $\epsilon$.
For example, for $\epsilon = 0.5$, Eve still gets the information
bit with probability $p \approx 0.55$, Bob's loss rate for the CTRL bits
is $R_\sCTRL \approx 0.83$, and his loss rate for the SWAP-$x$ bits
is $R_\sSWAPx \approx 0.73$.

\begin{table}[H]
\caption{The probability $p$ of Eve obtaining an information bit,
and the loss rates $R_\sCTRL$ and $R_\sSWAPx$
of CTRL and SWAP-$x$ bits (where $x \in \{01,10\}$),
respectively, for several values of the attack's parameter $\epsilon$.}
\label{table_epsilon_p_weaker}
\centering
\begin{tabular}{cccccccccccc}
\toprule
$\bm{\epsilon}$ & 0 & 0.1 & 0.2 & 0.3 & 0.4 & 0.5 & 0.6 & 0.7 & 0.8
& 0.9 & 1 \\
\midrule
$\mathbf{p}$ & 1 & 0.97 & 0.89 & 0.78 & 0.66 & 0.55 & 0.44 & 0.34
& 0.25 & 0.15 & 0 \\
$\mathbf{R_\sCTRL}$ & 1 & 0.99 & 0.97 & 0.94 & 0.89 & 0.83 & 0.76
& 0.67 & 0.57 & 0.46 & 0.33 \\
$\mathbf{R_\sSWAPx}$ & 0.83 & 0.83 & 0.82 & 0.79 & 0.76 & 0.73 & 0.68
& 0.63 & 0.58 & 0.53 & 0.5 \\
\bottomrule
\end{tabular}
\end{table}

For all values of $\epsilon$, the attack causes no errors.
However, in principle, it can be detected
because it causes different loss rates to different types of bits:
the loss rate experienced by Bob in the CTRL bits, $R_\sCTRL$,
is usually different from the loss rate in the SWAP-$x$ bits,
$R_\sSWAPx$ (see Table~\ref{table_epsilon_p_weaker} for details).
Therefore, in principle, the attack can be detected by a statistical
test for most values of $\epsilon$.

The loss rates become equal only for the value $\epsilon = \epsilon_0
\triangleq \sqrt{\frac{3 - \sqrt{3}}{2}} \approx 0.796$
(which gives $\kappa^2 = \frac{\epsilon^2}{3}$).
It seems that this specific attack \emph{cannot} be detected,
even in principle:
it causes no errors, and it causes the same loss rate for all qubits.
For this attack, Eve gets the information bit with probability
$p = \frac{1}{4}$, and the loss rates are
$R_\sCTRL = R_\sSWAPx = \frac{1}{\sqrt{3}} \approx 0.577$.
Therefore, this attack gives Eve a reasonable amount of information,
and it is not detectable by looking at errors or comparing loss rates.
(We can slightly modify the attack to make the loss rate the same
in both directions of the quantum channel, too.)

We conclude that this weaker attack gives Eve partial information,
causes no errors, and causes several loss rates.
We also conclude that since the loss rates caused by the attack
are usually different for different types of bits,
the attack can be detected, in principle,
for any value of $\epsilon$ except $\epsilon_0$.
However, for $\epsilon = \epsilon_0$, the attack seems undetectable.

\section{\label{sec_discuss}Discussion}
We have discussed a simpler and natural variant of the Mirror protocol
(the ``simplified Mirror protocol'') which is easier to implement.
We have found the simplified Mirror protocol to be completely
non-robust; therefore, this protocol is actually
an ``over-simplified'' Mirror protocol.
We have presented in Subsection~\ref{subsec_attack_full} an attack
giving Eve full information without causing any error;
in addition, since this attack also causes full loss of the CTRL bits,
we have presented in Subsection~\ref{subsec_attack_weaker}
weaker attacks giving Eve partial information, causing no errors,
and causing fewer losses. In particular, we have presented
a specific attack (characterized by the parameter $\epsilon = \epsilon_0
\triangleq \sqrt{\frac{3 - \sqrt{3}}{2}} \approx 0.796$)
that seems undetectable and gives Eve one quarter ($\frac{1}{4}$)
of all information bits.

Those attacks prove that the simplified Mirror protocol,
which allows Alice to use only three classical operations
(CTRL, SWAP-10, and SWAP-01), is completely non-robust.
On the other hand, the Mirror protocol is proved completely robust
(see Section~\ref{sec_mirror} and~\cite{mirror17}).
As explained in Section~\ref{sec_simplified},
the only difference between the simplified Mirror protocol and
the Mirror protocol is that the Mirror protocol allows
a fourth classical operation, SWAP-ALL;
therefore, allowing the SWAP-ALL operation is necessary for robustness.
More generally, the Mirror protocol probably
cannot be made much simpler while remaining robust:
its complexity is crucial for robustness.
Therefore, we have seen that if we want to use an SQKD protocol
that is experimentally feasible in a secure way,
we may have to use a relatively complicated protocol.

In this paper, we have not checked the experimental feasibility of
Eve's attacks, because Eve is usually assumed to be all-powerful.
Nonetheless, it can be interesting to check in the future
the experimental feasibility of those attacks and discover
whether the simplified Mirror protocol is flawed also in practice
and not ``only'' in theory. Other interesting directions for future
research include trying to find experimentally
feasible SQKD protocols that are simpler than the Mirror protocol,
and trying to find similar attacks against other QKD and SQKD protocols
that have no complete robustness proof.

\vspace{6pt} 

\authorcontributions{T.M. suggested to investigate the robustness of
the simplified protocol. M.B. suggested and designed the two attacks.
All authors performed the careful analysis of the attacks,
wrote the manuscript,
and reviewed and commented on the final manuscript.}

\funding{The work of Tal Mor and Rotem Liss was partly supported
by the Israeli MOD Research and Technology Unit.}

\acknowledgments{The authors thank Natan Tamari and Pavel Gurevich for
useful discussions about the experimental implementation of SWAP-ALL.}

\conflictsofinterest{The authors declare no conflict of interest.}

\abbreviations{The following abbreviations are used in this manuscript:\\

\noindent 
\begin{tabular}{@{}ll}
QKD & Quantum Key Distribution \\
SQKD & Semiquantum Key Distribution
\end{tabular}}

\reftitle{References}
\externalbibliography{yes}
\bibliography{mirror_attack}

\end{document}